\documentclass{aastex}          
\usepackage{spr-astr-addons}    

\begin{document}
%
\title{de Sitter expansion with anisotropic fluid in Bianchi type-I space-time}

\shorttitle{de Sitter expansion with anisotropic fluid in Bianchi type-I space-time}
\shortauthors{Akarsu \and K{\i}l{\i}n\c{c}}

\author{\"{O}zg\"{u}r Akarsu \altaffilmark{1}} \and \author{Can Battal K{\i}l{\i}n\c{c} \altaffilmark{2}}
\affil{Ege University, Faculty of Science, Dept. of Astronomy and Space Sciences, 35100 Bornova, {\.I}zmir/Turkey.}

\altaffiltext{1}{e-mail: ozgur.akarsu@mail.ege.edu.tr}
\altaffiltext{2}{e-mail: can.kilinc@ege.edu.tr}

\begin{abstract}
Some features of the Bianchi type-I universes in the presence of a fluid that wields an anisotropic equation of state (EoS) parameter are discussed in the context of general relativity. The models that exhibit de Sitter volumetric expansion due to the constant effective energy density (the sum of the energy density of the fluid and the anisotropy energy density) are of particular interest. We also introduce two locally rotationally symmetric models, which exhibit de Sitter volumetric expansion in the presence of a hypothetical fluid that has been obtained by minimally altering the conventional vacuum energy. In the first model, the directional EoS parameter on the $x$ axis is assumed to be -1, while the ones on the other axes and the energy density of the fluid are allowed to be functions of time. In the second model, the energy density of the fluid is assumed to be constant, while the directional EoS parameters are allowed to be functions of time.
\end{abstract}

\keywords{Bianchi type-I; de Sitter expansion; cosmological constant; anisotropic fluid}

%

%
%

\section{Introduction}
\label{intro}

Inhomogeneous and anisotropic universes have particularly been of interest in mathematical cosmology, rather than observational cosmology. This situation can be related to two main ingredients of modern cosmology: (i) The present universe had very well been described with Friedmann-Lemaitre models that are based on the spatially homogeneous and isotropic Robertson-Walker space-time. (ii) According to the inflationary paradigm the universe should have achieved an almost isotropic and homogeneous geometry at the end of the inflationary era (see \citealt{linde08} for a review of inflationary cosmology). However, for a realistic cosmological model one should consider spatially inhomogeneous and anisotropic space-times and then show whether they can evolve to the observed amount of homogeneity and isotropy. Because of the analytical difficulties in studying the inhomogeneous models, as a first step toward a more realistic model one may consider Bianchi type cosmological models, which form a large and almost complete class of relativistic cosmological models, which are homogeneous but not necessarily isotropic. The only spatially homogeneous but anisotropic models other than Bianchi type models are the Kantowski-Sachs locally symmetric family. See \cite{ellis99} for generalized, particularly anisotropic, cosmological models and \cite{ellis06} for a concise review on Bianchi type models.

In principle, once the metric is generalized to Bianchi types, the equation of state (EoS) parameter of the fluid can also be generalized in a way conveniently to wield anisotropy with the considered metric. In such models, where both the metric and EoS parameter of the fluid are allowed to exhibit an anisotropic character, the universe can exhibit non-trivial isotropization histories and it can be examined whether the metric and/or the EoS parameter of the fluid evolve toward isotropy. Thus, one gets the opportunity of constructing even more realistic models than the Bianchi type models where the fluid that fills the universe is assumed to be isotropic from the beginning.

Moreover, in recent years Bianchi universes have been gaining an increasing interest of observational cosmology, since the WMAP data \citep{hinshaw03,hinshaw07,hinshaw09} seem to require an addition to the standard cosmological model with positive cosmological constant that resembles the Bianchi morphology (see \citealt{jaffe05,jaffe06a,jaffe06b,campanelli06,campanelli07,hoftuft09}). According to this, the universe should have achieved a slightly anisotropic spatial geometry in spite of the inflation, contrary to generic inflationary models \citep{guth81,sato81,albrecht82,linde82,linde83,linde91,linde94} and that might be indicating a non-trivial isotropization history of the universe due to the presence of an anisotropic energy source. We may talk about two main classes of such models; according to whether this anisotropization occurs at an early time or at late times of the universe. The former class can be related with the inflaton field, which drives the inflation, while the latter one can be related with the dark energy (DE), which drives the late time acceleration of the universe.

In the context of the former class, the generic inflationary models can be modified in a way to end inflation with a slightly anisotropic spatial geometry; see, e.g., \cite{campanelli06,campanelli07}. A key ingredient of the inflationary models is an inflaton field, which generates an expansion like the one generated by the cosmological constant. In these models, space, and thus the inflaton field, is assumed to be homogeneous and isotropic from the beginning. The possibility of inflation in anisotropic space-time has first been investigated by \cite{barrow81}. Some authors have considered Bianchi type space-times in the presence of scalar fields, e.g., \cite{feinstein93,aguirregabiria93}. \cite{ford89} discussed the possibility of inflationary models in which inflation is driven by a vector field, which gives rise to an anisotropic EoS parameter, rather than a scalar field, for the first time. However, it was suffering from the fine-tuning problem. Recently, \cite{koivisto08c} have considered several new classes of viable vector field as alternatives to the inflaton within in the Bianchi type-I framework. \cite{golovnev08} have constructed a successful model, which either could give a completely isotropic universe or slightly anisotropic universe at the end of the inflation.

In the context of the latter class, the isotropy of space that has been achieved in the inflationary era can be distorted during the contemporary acceleration of the universe by modifying the DE in a way to wield anisotropic pressure, see, e.g., \cite{koivisto08a,koivisto08b,koivisto08c,rodrigues08}. It is well known that the inclusion of a positive cosmological constant, which is mathematically equivalent to the vacuum energy with $p=-\rho$ and which is the simplest candidate for the DE, may lead to the explanation of this observed acceleration of the universe.

Note that the vacuum energy (the cosmological constant) is coming into question in both early and late time acceleration of the universe. The solution of Einstein's field equations in the presence of a positive cosmological constant for a homogeneous and isotropic model results in an exponentially expanding universe, which is known as the de Sitter model \citep{gronhervik}. Thus, de Sitter and de Sitter like expansions, hence the cosmological constant and its alterations, are of particular interest for contemporary cosmology.

\cite{wald83} investigated the asymptotic behavior of initially expanding homogeneous cosmological models with a positive cosmological constant and showed that such models, of all Bianchi types (only type-IX under some conditions), exponentially evolve toward the de Sitter solution. \cite{gron85} investigated the expansion anisotropy during the inflationary era for a vacuum Bianchi type-I universe with a non-vanishing positive cosmological constant, and discussed the anisotropic generalization of the de Sitter solution. \cite{beesham94} showed that Bianchi type-I models in the presence of a positive cosmological constant cannot be of the pure de Sitter type, unless the gravitational "constant" takes negative values. \cite{kalligas94} showed that de Sitter inflation is allowed at least at late times of the universe within the Bianchi type-I framework. \cite{arbab97} constructed a Bianchi type-I cosmological model, in the presence of variable $G$ and $\Lambda$ and bulk viscosity, that exhibits de Sitter expansion. \cite{kumar07} gave an exponentially expanding Bianchi type-I model in the presence of a hypothetical perfect fluid. \cite{akarsu10} presented a locally rotationally symmetric (LRS) Bianchi type-I model that exhibits de Sitter volumetric expansion in a mixture of a perfect fluid and a fluid that wields a special form of a dynamical and anisotropic EoS parameter.

\cite{barrow97} presented an analysis of the cosmological evolution of matter sources that possess small anisotropic pressures (electric and magnetic fields, collisionless relativistic particles, gravitons, anti-symmetric axion fields in low-energy string cosmologies, spatial curvature anisotropies and stresses arising from simple topological defects), and he discussed the effects of inflation on the anisotropy of the pressures.

Considering the above discussion, one may first think to distort the isotropy of the vacuum energy so as to obtain anisotropy in the geometry of the universe in general relativity. Thus, in this study, Bianchi type-I models that exhibit de Sitter volumetric expansion in the presence of a vacuum energy that has minimally been altered in a way to wield anisotropic EoS parameter will be of our particular interest. We first, in Sect. 2, discuss some features of the Bianchi type-I universe filled with a fluid that wields an anisotropic EoS parameter. We define the effective energy density as the sum of the energy density of the fluid and the anisotropy energy density, and specialize the discussion to the models that exhibit de Sitter volumetric expansion due to the constant effective energy density. We then, in Sect. 3, examine the solutions for a LRS Bianchi type-I space-time, where the universe exhibits de Sitter volumetric expansion in the presence of a hypothetical fluid, that has been obtained by minimally altering the conventional vacuum energy, and we present two exact models. In the first model, in Sect. 3.1, the directional EoS parameter on the $x$ axis is assumed to be -1, while the ones on the other axes and the energy density of the fluid are allowed to be functions of time. In the second model, in Sect. 3.2, the energy density of the fluid is assumed to be constant, while the directional EoS parameters are allowed to be functions of time.

\section{A general discussion on the dynamics of the models in the presence of an anisotropic fluid}
\label{sec:1}
\subsection{The field equations in the presence of anisotropic fluid}
We consider the diagonal metric of the spatially flat, homogeneous but anisotropic Bianchi type-I space-time,
\begin{equation}
ds^{2}=dt^{2}-A(t)^{2}dx^{2}-B(t)^{2}dy^{2}-C(t)^{2}dz^{2},
\end{equation}
where $A(t)$, $B(t)$ and $C(t)$ are the directional scale factors, functions of the cosmic time, $t$. 

Within the framework of the metric given by (1), the energy-momentum tensor of a fluid can be written, most generally, in anisotropic diagonal form as follows:
\begin{equation}
{{T}_{\nu}}^{\mu}=\mathrm{diag}[{{T}_{0}}^{0},{{T}_{1}}^{1},{{T}_{2}}^{2},{{T}_{3}}^{3}].
\end{equation}

The Einstein field equations, in natural units ($8\pi G=1$ and $c=1$), are
\begin{equation}
R_{\mu\nu}-\frac{1}{2}Rg_{\mu\nu}=-{T}_{\mu\nu},
\end{equation}
where $g_{\mu\nu}$ is the metric tensor; $g_{\mu\nu}u^{\mu}u^{\nu}=1$ ($u^{\mu}=(1,0,0,0)$ is the four-velocity vector); $R_{\mu\nu}$ is the Ricci tensor; $R$ is the Ricci scalar, ${T}_{\mu\nu}$ is the energy-momentum tensor.

In a comoving coordinate system, the Einstein field equations (3), for the Bianchi type-I space-time (1) in case of the anisotropic energy-momentum tensor given by (2), read 
\begin{equation}
\frac{\dot{A}}{A}\frac{\dot{B}}{B}+\frac{\dot{A}}{A}\frac{\dot{C}}{C}+\frac{\dot{B}}{B}\frac{\dot{C}}{C}={{T}_{0}}^{0},
\end{equation}
\begin{equation}
\frac{\ddot{B}}{B}+\frac{\ddot{C}}{C}+\frac{\dot{B}}{B}\frac{\dot{C}}{C}={{T}_{1}}^{1},
\end{equation}
\begin{equation}
\frac{\ddot{A}}{A}+\frac{\ddot{C}}{C}+\frac{\dot{A}}{A}\frac{\dot{C}}{C}={{T}_{2}}^{2},
\end{equation}
\begin{equation}
\frac{\ddot{A}}{A}+\frac{\ddot{B}}{B}+\frac{\dot{A}}{A}\frac{\dot{B}}{B}={{T}_{3}}^{3},
\end{equation}
where an overdot denotes $d/dt$.

The directional Hubble parameters, which determine the expansion rates of the universe in the directions of $x$, $y$ and $z$, for the metric given in (1) can be defined as 
\begin{equation}
H_{x}\equiv\frac{\dot{A}}{A}\textnormal{,}\qquad H_{y}\equiv\frac{\dot{B}}{B}\qquad \textnormal{and}\qquad H_{z}\equiv\frac{\dot{C}}{C},
\end{equation}
respectively, and then the mean Hubble parameter, which determines the volumetric expansion rate of the universe, can be given as
\begin{equation}
H=\frac{1}{3}\frac{\dot{V}}{V}=\frac{H_{x}+H_{y}+H_{z}}{3},
\end{equation}
where $V=ABC$ is the volume of the universe.

\subsection{The anisotropy of the expansion}

The anisotropy of the expansion can be parametrized by using the directional Hubble parameters (8) and the mean Hubble parameter (9) of the expansion,
\begin{equation}
\Delta\equiv\frac{1}{3}\sum_{i=1}^{3}\left(\frac{H_{i}-H}{H}\right)^{2},
\end{equation}
where $H_{i}$ (i=1,2,3) represents the directional Hubble parameters in the directions of $x$, $y$ and $z$, respectively. $\Delta=0$ corresponds to isotropic expansion. The Bianchi type-I universe approaches spatial isotropy if $\Delta\rightarrow 0$, $V\rightarrow +\infty$ and ${T}^{00}>0$ as $t\rightarrow+\infty$ (see \citealt{collins73} for details).

The anisotropy parameter of the expansion can also be given in terms of the components of the energy-momentum tensor (2) and the mean Hubble parameter (9) by using the evolution equations (5)-(7);
\begin{equation}
\Delta=\frac{1}{9H^{2}}\sum_{i,j=1}^{3}\left[\lambda_{j}+\int ({{T}_{i}}^{i}-{{T}_{j}}^{j})Vdt \right]^{2}V^{-2}\; \textnormal{and} \; i>j,
\end{equation}
where the $\lambda_{j}$ are real constants. One can immediately see that the above equation is reduced to the one given by \cite{gron85}: 
\begin{equation}
\Delta=\frac{{\lambda_{1}}^{2}+{\lambda_{2}}^{2}+{\lambda_{3}}^{2}}{9H^{2}}V^{-2}
\end{equation}
for an isotropic fluid, i.e., where ${{T}_{1}}^{1}={{T}_{2}}^{2}={{T}_{3}}^{3}$.

\subsection{Anisotropy energy density in the presence of an anisotropic fluid}

The energy density of the fluid $\rho$, which corresponds to ${{T}_{0}}^{0}$, can immediately be written in terms of the mean Hubble parameter $H$ and the anisotropy parameter of the expansion $\Delta$ by using the constraint equation (4), 
\begin{equation}
\rho={{T}_{0}}^{0}=3H^{2}\left(1-\frac{\Delta}{2}\right).
\end{equation}
This is equivalent to the generalized Friedmann equation for a Bianchi type-I space-time (see \citealt{ellis99} and \citealt{barrow95} for the generalized Friedmann equation). According to this, for a given value of the mean Hubble parameter $H$ in Bianchi type-I space-time, the anisotropy of the expansion lowers down the energy density; the highest energy density is achieved in case of isotropic expansion (i.e., $\Delta=0$) and the anisotropy of the expansion cannot be arbitrary because the condition to observe $\rho>0$ for a comoving observer is $\Delta<2$.

By moving from (13), the anisotropy energy density (namely, the energy density that associated with the anisotropy of the expansion) may be defined as follows:
\begin{equation}
\rho_{\beta}\equiv\frac{3}{2}H^{2}\Delta.
\end{equation}
Then, using (11) in this definition, we obtain
\begin{equation}
\rho_{\beta}=\frac{1}{6}\sum_{i,j=1}^{3}\left[\lambda_{j}+\int ({{T}_{i}}^{i}-{{T}_{j}}^{j})Vdt \right]^{2}V^{-2}\quad \textnormal{and} \quad i>j,
\end{equation}
for the anisotropy energy density. One can observe that the anisotropy of the fluid also contributes to the determination of the $\rho_{\beta}$ via the integral term. In case of an isotropic fluid the integral term vanishes and $\rho_{\beta}$ is reduced to the following form:
\begin{equation}
\rho_{\beta}=\frac{1}{6}\left({\lambda_{1}}^{2}+{\lambda_{2}}^{2}+{\lambda_{3}}^{2}\right)V^{-2},
\end{equation}
which is equivalent to the one defined by \cite{barrow81} in case of an isotropic fluid. According to this equation, (16), $\rho_{\beta}$ decreases monotonically as the volume of the universe increases and converges to null as $V\rightarrow+\infty$, while it diverges as $V\rightarrow 0$. On the other hand, as can be seen from (15), $\rho_{\beta}$ can exhibit non-trivial behaviors if the fluid is allowed to be anisotropic. In other words, in the presence of an anisotropic fluid, we may have cosmological models with non-trivial isotropization histories, e.g., models in which $\rho_{\beta}$ does not diverge as $V\rightarrow 0$ and/or $V\rightarrow+\infty$.

\subsection{Conservation laws for the anisotropic fluid and vacuum energy}

Allowing for anisotropy in the pressure of the fluid, and thus in its EoS parameter, gives rise to new possibilities for the evolution of the energy source. To see this, we first parametrize the energy-momentum tensor given in (2) as follows:
\begin{eqnarray}
{{T}_{\nu}}^{\mu}=\mathrm{diag}[\rho,-{p_{x}},-{p_{y}},-{p_{z}}]\\
\nonumber =\mathrm{diag}[1,-{w_{x}},-{w_{y}},-{w_{z}}]\rho\\
\nonumber =\mathrm{diag}[1,-w,-(w+\gamma),-(w+\delta)]\rho,
\end{eqnarray}
where ${p_{x}}$, ${p_{y}}$ and ${p_{z}}$ are the pressures and ${w_{x}}$, ${w_{y}}$ and ${w_{z}}$ are the directional EoS parameters on the $x$, $y$ and $z$ axes, respectively; $w$ is the deviation-free EoS parameter of the fluid. The deviation from isotropy is parametrized by setting ${w_{x}}=w$ and then introducing skewness parameters $\delta$ and $\gamma$, which are the deviations from $w$, respectively, on the $y$ and $z$ axes. $w$, $\delta$ and $\gamma$ are not necessarily constants and might be functions of the cosmic time, $t$.

The conservation of the energy-momentum tensor of the fluid parametrized in (17), i.e., ${{T}^{\mu\nu}}_{;\nu}=0$, leads to the following equation:
\begin{equation}
\dot{\rho}+3(1+w)\rho H+\delta\rho H_{y}+\gamma\rho H_{z}=0,
\end{equation}
where the last two terms, the terms with $\delta$ and $\gamma$, arise due to the anisotropy of the fluid. Now we can briefly examine what possibilities arise due to these two terms when the conventional vacuum energy (which is mathematically equivalent to the cosmological constant $\Lambda$ and which can be represented with an EoS parameter in the form of $p=-\rho$) is minimally altered in a way to wield an anisotropic EoS parameter. Obviously, when we consider the conventional vacuum energy, which is isotropic (i.e., $\gamma=\delta=0$), if $w=-1$, then the energy density ($\rho$) is necessarily constant and vice versa. However, this is not the case when the energy-momentum tensor of the fluid is generalized to the form given in (17), namely, when an anisotropic EoS parameter is allowed for the fluid. In that case, if the energy density is assumed to be constant, $\rho=const.$, (18) is reduced to the following equation:
\begin{equation}
3(1+w)H+\delta H_{y}+\gamma H_{z}=0,
\end{equation}
thus, $w$ is not necessarily constant. Similarly, if $w=-1$, (18) is reduced to the following equation:
\begin{equation}
\dot{\rho}+\delta\rho H_{y}+\gamma\rho H_{z}=0,
\end{equation}
thus, the energy density is not necessarily constant either.

\subsection{Generalized Friedmann equation and de Sitter expansion}

Using the energy-momentum tensor parametrized in (17), the anisotropy energy density can be written in terms of the directional EoS parameters as follows:
\begin{equation}
\rho_{\beta}=\frac{1}{6}\sum_{i,j=1}^{3}\left[\lambda_{j}+\int ({{w}_{j}}-{{w}_{i}})\rho Vdt \right]^{2}V^{-2}\; \textnormal{and} \; i>j.
\end{equation}
Note that the behavior of the anisotropy energy density is not only determined by the $\lambda_{j}$ constants and the volume of the universe as in (16), but also by the directional EoS parameters ($w_{x}$, $w_{y}$ and $w_{z}$) and the energy density of the fluid ($\rho$) via the integral term.

We may write down the generalized Friedmann equation by considering (13) and (14) and define the effective energy density $\rho_{\textnormal{ef}}$, which determines the volumetric expansion rate of the universe,
\begin{equation}
3H^{2}=\rho+\rho_{\beta}\equiv\rho_{\textnormal{ef}}.
\end{equation}
From the definition of the mean Hubble parameter given in (9) the volume of the universe can be obtained as follows:
\begin{equation}
V=c_{1}e^{3\int{H}dt},
\end{equation}
where $c_{1}$ is a positive constant of integration. Thus, using (22) in (23) the volume of the universe can be obtained in terms of the effective energy density,
\begin{equation}
V=c_{1}e^{\sqrt{3}\int{\sqrt{\rho_{\textnormal{ef}}}}dt}.
\end{equation}
According to this, the condition for the de Sitter volumetric expansion does not correspond to a constant energy density of the fluid but to a constant effective energy density, i.e., $\rho_{\textnormal{ef}}=\rho+\rho_{\beta}=const.$, which leads to
\begin{eqnarray}
V=c_{1}e^{\sqrt{3\rho_{\textnormal{ef}}}t}.
\end{eqnarray}
Obviously, the universe exhibits de Sitter expansion when $\rho_{\beta}$ is null in the presence of a positive cosmological constant, i.e., in the presence of conventional vacuum energy that wields a constant energy density. However, if $\rho_{\beta}$ is not null, then $\rho_{\textnormal{ef}}$ is not constant, but a decreasing function of the cosmic $t$ as long as the universe expands, since $\rho_{\beta}\propto V^{-2}$. Thus, as also shown by \cite{beesham94}, the behavior of Bianchi type-I inflationary solutions cannot be of the pure de Sitter type, even when $p=-\rho$, but are of power-law type.

What if we consider the anisotropy energy density together with a mixture of the conventional vacuum energy and perfect fluids? The conventional perfect fluids, i.e., radiation, pressureless matter etc. can be described by an EoS parameter in the form of $p=w\rho$ and their energy densities change as $V^{-(1+w)}$. Thus, the cosmological constant will be dominant as $V\rightarrow+\infty$, provided that $w>-1$, and according to the cosmic no-hair theorem for Einstein gravity introduced by \cite{wald83}, the universe will exponentially evolve toward the de Sitter universe. In other words, Bianchi type-I models in the presence of a positive cosmological constant isotropize and their volumetric expansion rates approach de Sitter expansion as $V\rightarrow+\infty$. On the other hand, according to (16), no matter how small the anisotropy energy density is, compared with the other sources, $\rho_{\beta}$ will eventually dominate any perfect fluid and govern the dynamics of the expansion in the very early evolution of the universe as $V\rightarrow 0$, provided that $w<1$, i.e., the universe will approximate the Kasner vacuum solution. However, these cases are not implied once the implicitly assumed isotropy of the vacuum energy is relaxed. This is because, according to (21), $\rho_{\beta}$ does not have to be proportional with $V^{-2}$ in case of an anisotropic fluid, and it may not diverge as $V\rightarrow 0$ and/or $V\rightarrow+\infty$.

Our final remark in this section concerns the condition for isotropization of a Bianchi type-I universe that exhibits de Sitter volumetric expansion. The condition for isotropization mentioned in Sect. 2.3 can be rewritten as follows: $\rho_{\beta}\rightarrow0$ as $t\rightarrow+\infty$ since $H$ is constant, and this leads to $\dot{\rho_{\beta}}<0$ as $t\rightarrow+\infty$, since $\rho_{\beta}>0$ by definition. This, in addition, leads to $\dot{\rho}>0$ as $t\rightarrow+\infty$ from $\dot{\rho_{\beta}}+\dot{\rho}=0$. Thus, the condition for isotropization in the Bianchi type-I models that exhibit de Sitter volumetric expansion imposes on the fluid the constraint to behave like a phantom energy, i.e., to exhibit an increasing energy density as $V$ increases. On the other hand, if the energy density of the fluid is constant ($\rho=const.$), then the anisotropy energy density is also constant ($\rho_{\beta}=const.$). If $\dot{\rho}<0$, then  $\dot{\rho_{\beta}}>0$; thus for models in which the energy density of the fluid is decreasing, the anisotropy energy density (thus, the anisotropy of the expansion) increases as $t$ increases.

\section{Exponentially expanding LRS models in the presence of anisotropic fluid}
\label{sec:3}
In the following, we examine the cosmological models that exhibit de Sitter volumetric expansion within the LRS Bianchi type-I framework and present two exact models.

The LRS Bianchi type-I metric reads
\begin{equation}
ds^{2}=dt^{2}-A(t)^{2}dx^{2}-B(t)^{2}(dy^{2}+dz^{2}).
\end{equation}
Thus, $H_{y}=H_{z}$, and in the following they are represented by $H_{y,z}$. The energy-momentum tensor given in (17) can be customized for the LRS Bianchi-I metric by choosing $\delta=\gamma$,
\begin{eqnarray}
{{T}_{\nu}}^{\mu}=\mathrm{diag}[1,-w,-(w+\gamma),-(w+\gamma)]\rho.
\end{eqnarray}
Considering (26) and (27), the Einstein field equations (4)-(7) can be reduced to the following system of equations:
\begin{equation}
\frac{\dot{B}^{2}}{B^{2}}+2\frac{\dot{A}}{A}\frac{\dot{B}}{B}=\rho,
\end{equation}
\begin{equation}
\frac{\dot{B}^{2}}{B^{2}}+2\frac{\ddot{B}}{B}=-w\rho,
\end{equation}
\begin{equation}
\frac{\ddot{B}}{B}+\frac{\dot{B}}{B}\frac{\dot{A}}{A}+\frac{\ddot{A}}{A}=-(w+\gamma)\rho.
\end{equation}
Then we have initially five variables ($A$, $B$, $\rho$, $w$, $\gamma$) and three linearly independent equations, namely the three Einstein field equations (28)-(30). Thus we will need two additional constraints to close the system of equations. However, before introducing the constraints to close the system of equations, we can give $\rho$, $w$ and $\gamma$ in terms of the mean Hubble parameter and the directional Hubble parameter on the $x$ axis, manipulating the field equations (28)-(30), 
\begin{equation}
\rho=3H^{2}-\frac{3}{4}\left(H_{x}-H\right)^2,
\end{equation}
\begin{equation}
w=\frac{\frac{d}{dt}(H_{x}-3H)-\frac{3}{4}(H_{x}-3H)^{2}}{3H^{2}-\frac{3}{4}\left(H_{x}-H\right)^2},
\end{equation}
\begin{equation}
\gamma=\frac{\frac{3}{2}\frac{d}{dt} (H-H_{x})+\frac{9}{2}(H-H_{x})H}{3H^{2}-\frac{3}{4}\left(H_{x}-H\right)^2}.
\end{equation}
The anisotropy energy density can also be given in terms of $H$ and $H_{x}$ by using (13), (14) and (28),
\begin{equation}
\rho_{\beta}=\frac{3}{4}(H_{x}-H)^{2}.
\end{equation}
One may check that the summation of (31) and (34) leads to $\rho_{\textnormal{ef}}=\rho+\rho_{\beta}=3H^{2}$.

As the first constraint to close the system of equations, the effective energy density is assumed to be constant throughout the history of the universe:
\begin{equation}
\rho_{\textnormal{ef}}=3k^{2},
\end{equation}
where $k$ is a positive constant and thus from (22)
\begin{equation}
H=k,
\end{equation}
which corresponds to the well-known de Sitter volumetric expansion, i.e., 
\begin{equation}
V=AB^{2}=c_{1}e^{3kt}.
\end{equation}
The solution of (33) by considering (31) and (36) gives the following equation for the evolution of the directional scale factor on the $x$ axis:
\begin{equation}
A=\kappa e^{kt+\frac{\lambda}{3k}e^{-3kt}+\frac{2}{9k}\left(e^{-3kt}\int{e^{3kt}\Gamma(t){dt}}-\int{\Gamma(t){dt}}\right)},
\end{equation}
where $\kappa>0$ and $\lambda$ are real constants and $\Gamma(t)=\gamma\rho$ is the skewness of the pressure. Using (38) in (37) the directional scale factor on the $y$ and $z$ axes is obtained as follows:
\begin{equation}
B=\left({\frac{c_{1}}{\kappa}}\right)^{1/2}e^{kt-\frac{\lambda}{6k}e^{-3kt}-\frac{1}{9k}\left(e^{-3kt}\int{e^{3kt}\Gamma(t){dt}}-\int{\Gamma(t){dt}}\right)}.
\end{equation}
Using the scale factors (38) and (39), the directional Hubble parameters are obtained as follows:
\begin{equation}
H_{x}=k-\lambda e^{-3kt}-\frac{2}{3}e^{-3kt}\int{\Gamma(t) e^{3kt}}dt,
\end{equation}
\begin{equation}
H_{y,z}=k+\frac{\lambda}{2}e^{-3kt}+\frac{1}{3}e^{-3kt}\int{\Gamma(t) e^{3kt}}dt.
\end{equation}
Using (36), (40) and (41) in (10) the anisotropy of the expansion, $\Delta$, and thus the anisotropy energy density, $\rho_{\beta}$, are obtained as
\begin{equation}
\Delta=\frac{2}{3}\frac{\rho_{\beta}}{k^{2}}=\frac{1}{18}\frac {{e^{-6kt}} \left( 3{\lambda}+2\int{e^{3kt}}\Gamma(t){dt}\right)^{2}}{{k}^{2}}.
\end{equation}
It can be seen that the skewness of the pressure $\Gamma(t)$ also contributes to the evolution of the cosmological parameters.

One can immediately obtain the cosmological parameters once a function for $\Gamma(t)$ is chosen. Alternatively, one can obtain the parameters by introducing one more constraint on one of the cosmological parameters; for instance, on the energy density of the fluid $\rho$ or on the EoS parameter. Trivially, one may also assume that the fluid is a perfect fluid (i.e., $\gamma=0$; thus $\Gamma$ is null) and would then obtain the solutions given by \cite{kumar07} for an exponentially expanding Bianchi type-I universe. One can observe that for any perfect fluid, i.e., in case of $\Gamma=0$, the initial anisotropy of the expansion dies away monotonically as $t$ increases, while it may exhibit various non-trivial behaviors according to the function chosen for $\Gamma(t)$.

\subsection{Model for $\rho_{\textnormal{ef}}=const.$ and $w=-1$}
\label{subsec:1}
Note that we need one more constraint in addition to the one given in (35) to fully determine the system. Since the first assumption given by (35) causes de Sitter volumetric expansion, it seems reasonable to choose our last assumption as
\begin{equation}
w=-1.
\end{equation}
If we solve the system of equations which consists of the three Einstein field equations (28)-(30) and the two constraints given by (35) and (43), we obtain the following exact expressions for the scale factors:
\begin{equation}
A={c_{1}}e^{kt}\left(\kappa k^{-1}+3\lambda e^{-3kt}\right)^{-2/3},
\end{equation}
\begin{equation}
B=e^{kt}\left(\kappa k^{-1}+3\lambda e^{-3kt}\right)^{1/3}.
\end{equation}
The directional Hubble parameters are obtained as follows:
\begin{equation}
H_{x}=k+\frac{6\lambda k^{2}}{\kappa e^{3kt}+3\lambda k},
\end{equation}
\begin{equation}
H_{y,z}=k-\frac{3\lambda k^{2}}{\kappa e^{3kt}+3\lambda k}.
\end{equation}
The anisotropy parameter of the expansion and thus the anisotropy energy density are found to be dynamical,
\begin{equation}
\Delta =\frac{2}{3}\frac{\rho_{\beta}}{k^{2}}=\frac{18\lambda^{2}k^{2}}{\left(\kappa e^{3kt}+3\lambda k\right)^{2}}.
\end{equation}
The energy density of the fluid is also found to be dynamical in a way so as to secure $\rho+\rho_{\beta}=3k^{2}$,
\begin{equation}
\rho=3k^{2}\left(1-\frac{9\lambda^{2}k^{2}}{\left(\kappa e^{3kt}+3\lambda k\right)^{2}}\right)
\end{equation}
and the skewness parameter of the EoS parameter is obtained as follows:
\begin{equation}
\gamma =-\frac{27\lambda^{2}k^{2}}{\kappa e^{3kt}\left(\kappa e^{3kt}+6\lambda k\right)}.
\end{equation}

One can check that this solution yields (20). The positiveness condition on the energy density of the fluid ($\rho>0$) imposes the restriction that the anisotropy of the expansion is smaller than 2 ($\Delta<2$) and that imposes $\lambda$ to be positive. Thus, only those models with positive $\lambda$ values are viable, and hence in the following we will consider only the positive $\lambda$ values.

$\Delta$ decreases monotonically as $t$ increases and $\Delta\rightarrow 0$ as $t\rightarrow+\infty$. Thus, the space approaches isotropy as $t\rightarrow+\infty$ in this model. On the other hand, $\Delta\rightarrow 2\left[1-\kappa/(\kappa+3\lambda k)\right]^{2}$ as $t\rightarrow 0$.

$\rho\rightarrow 3k^{2}$ and $\gamma\rightarrow 0$ as $t\rightarrow+\infty$. That is, the EoS parameter of the fluid isotropizes as $t$ increases and mimics the cosmological constant as $t\rightarrow+\infty$. On the other hand, $\rho\rightarrow 3k^{2}\left({1-\left(1+\frac{\kappa}{\lambda k}\right)^{-2}}\right)$ and $\gamma\rightarrow -\frac{27{\lambda}^{2}k^{2}}{{\kappa}^{2}+6\lambda\kappa k}$, which is always a negative number, as $t\rightarrow 0$. Therefore, as expected from the isotropization condition discussed in Sect. 2.5., $\rho$ behaves like a phantom energy and increases as $V$ increases. This result is consistent with the following situation: $\gamma$ is always negative and thus EoS parameters on the $y$ and $z$ axes are passing the phantom divide line, i.e., $w_{y,z}<-1$.

\subsection{Model for $\rho_{\textnormal{ef}}=const.$ and $\rho=const.$}
\label{subsec:2}
While $w$ is assumed to be $-1$ so as to close the system in the preceding section, it is allowed to be a function of the cosmic time $t$ in this section, but, this time, the energy density of the fluid is assumed to be constant:
\begin{equation}
\rho(t)=const.=\rho
\end{equation}
If we solve the system of equations which consists of the three Einstein field equations (28)-(30) and the two constraints given by (35) and (51), we obtain the following exact expressions for the scale factors:
\begin{equation}
A=\kappa^{-2}{c_{1}}e^{kt\pm \frac{2}{3}{\sqrt{9k^2-3\rho}}\,t},
\end{equation}
\begin{equation}
B=\kappa e^{kt\mp \frac{1}{3}{\sqrt{9k^2-3\rho}}\,t}.
\end{equation}
The directional Hubble parameters are obtained as follows:
\begin{equation}
H_{x}=k\pm \frac{2}{3}{\sqrt{9k^2-3\rho}},
\end{equation}
\begin{equation}
H_{y,z}=k\mp \frac{1}{3}{\sqrt{9k^2-3\rho}}.
\end{equation}
The anisotropy parameter of the expansion and thus the anisotropy energy density are found to be non-dynamical,
\begin{equation}
\Delta=\frac{2}{3}\frac{\rho_{\beta}}{k^{2}}=2-\frac{2}{3}\frac{\rho}{k^{2}}.
\end{equation}
The deviation-free EoS parameter of the fluid is obtained as follows:
\begin{equation}
w=\frac{\rho-6k^{2}}{\rho}\pm 2\frac{k}{\rho}\sqrt{9k^{2}-3\rho}=-1-\frac{\Delta\mp\sqrt{2\Delta}}{1-\frac{\Delta}{2}}.
\end{equation}
The skewness parameter of the EoS parameter is obtained as follows:
\begin{equation}
\gamma =\mp 3\frac{k}{\rho}\sqrt{9k^{2}-3\rho}=\mp\frac{3\sqrt{\frac{\Delta}{2}}}{1-\frac{\Delta}{2}}.
\end{equation}

One can check that this solution yields (19). The directional Hubble parameters, the anisotropy of the expansion (thus, the anisotropy energy density) and the directional EoS parameters are all constants throughout the history of the universe.

It is obvious that $\rho\leq3k^{2}$, thus $\rho_{\beta}\leq3k^{2}$ and $\Delta\leq2$. We recover the conventional vacuum energy and isotropic expansion when $\rho=3k^{2}$, because $w=-1$, $\gamma=0$ and $\Delta=0$ in that case. However, when $\rho<3k^{2}$, both the expansion and the fluid deviate from isotropy, because $w>-1$, $\gamma<0$ or $w<-1$, $\gamma>0$  and $\Delta>0$ in this case. It can be observed that, while the EoS parameter on the $x$ axis is in the quintessence region ($w>-1$), the ones on the $y$ and $z$ axes are in the phantom region ($w+\gamma<-1$), or vice versa. According to this, while the expansion of the $x$ axis acts so as to decrease the energy density of the fluid, the expansion of the $yz$ plane acts so as to increase the energy density of the fluid or vice versa. However, in total, the decrements and increments compensate for each other and the energy density of the fluid does not change.

\section{Conclusion}
\label{sec:4}

On motivating from the increasing evidence for the need of a geometry that resembles Bianchi morphology to explain the observed anisotropy in the WMAP data, we have discussed some features of the Bianchi type-I universes in the presence of a fluid that wields an anisotropic equation of state (EoS) parameter in general relativity. We have focused on those models that exhibit de Sitter volumetric expansion in the presence of a hypothetical fluid obtained by distorting the EoS parameter of the conventional vacuum energy in a way so as to wield anisotropy.

We have also given two exact solutions within the locally rotationally symmetric Bianchi type-I framework. In both models the effective energy density (the sum of the energy density of the fluid and the anisotropy energy density) has been assumed to be constant so as to secure the de Sitter volumetric expansion.

In the first model, the directional EoS parameter on the $x$ axis has been assumed to be -1. The anisotropy of the expansion, the energy density and the anisotropy of the fluid have been found to be dynamical. While the anisotropy of the expansion and the anisotropy of the fluid decrease and tend to null as the universe expands, the energy density of the fluid increases and approaches to its maximum value. The fluid approximates the conventional vacuum energy as the universe evolves.

In the second model, the energy density of the fluid has been assumed to be constant. The anisotropy of the expansion and the anisotropy of the fluid have been found to be non-dynamical. When the maximum value of the energy density of the fluid ($3k^{2}$, where $k$ is a positive constant) is considered, the universe expands isotropically and the fluid mimics the conventional vacuum energy. Lower values of the energy density of the fluid give rise to an anisotropy both in the expansion and EoS parameter of the fluid.

%
%

%

%
%

%

%
\acknowledgments
\"{O}zg\"{u}r Akarsu was supported in part by The Scientific and Technological Research Council of Turkey (T\"{U}B{\.I}TAK). Some of this work was carried out while \"{O}. Akarsu was visiting the Department of Applied Mathematics and Theoretical Physics (DAMTP), University of Cambridge. \"{O}. Akarsu would also like to thank Jonathan Middleton for the discussions he had with him.


%

%

\end{document}